\documentclass[aps,prl,onecolumn,amsmath,amssymb,superscriptaddress]{revtex4}
\usepackage{graphicx}
\usepackage{amssymb}
\usepackage{natbib}

\begin{document}

\title{Robust Upward Dispersion of the Neutron Spin Resonance in the Heavy Fermion Superconductor Ce$_{1-x}$Yb$_{x}$CoIn$_5$}

\author{Yu Song}
\affiliation{ Department of Physics and Astronomy,
Rice University, Houston, Texas 77005, USA }

\author{John Van Dyke}
\affiliation{ Department of Physics,
University of Illinois at Chicago, Chicago, Illinois 60607, USA}

\author{I. K. Lum}
\affiliation{Materials Science and Engineering Program, University of California, San Diego, La Jolla, California 92093, USA}
\affiliation{Department of Physics, University of California, San Diego, La Jolla, California 92093, USA}
\affiliation{Center for Advanced Nanoscience, University of California, San Diego, La Jolla, California 92093, USA}

\author{B. D. White}
\affiliation{Department of Physics, University of California, San Diego, La Jolla, California 92093, USA}
\affiliation{Center for Advanced Nanoscience, University of California, San Diego, La Jolla, California 92093, USA}

\author{Sooyoung Jang}
\affiliation{Materials Science and Engineering Program, University of California, San Diego, La Jolla, California 92093, USA}
\affiliation{Department of Physics, University of California, San Diego, La Jolla, California 92093, USA}
\affiliation{Center for Advanced Nanoscience, University of California, San Diego, La Jolla, California 92093, USA}

\author{Duygu Yazici}
\affiliation{Materials Science and Engineering Program, University of California, San Diego, La Jolla, California 92093, USA}
\affiliation{Department of Physics, University of California, San Diego, La Jolla, California 92093, USA}
\affiliation{Center for Advanced Nanoscience, University of California, San Diego, La Jolla, California 92093, USA}

\author{L. Shu}
\affiliation{State Key Laboratory of Surface Physics, Department of Physics, Fudan University, Shanghai 200433, China}
\affiliation{Collaborative Innovation Center of Advanced Microstructures, Fudan University, Shanghai 200433, China}

\author{A. Schneidewind}
\affiliation{J\"{u}lich Center for Neutron Science JCNS, Forschungszentrum J\"{u}lich GmbH, Outstation at MLZ, D-85747, Garching, Germany}

\author{Petr $\rm \check{C}$ermak}
\affiliation{J\"{u}lich Center for Neutron Science JCNS, Forschungszentrum J\"{u}lich GmbH, Outstation at MLZ, D-85747, Garching, Germany}

\author{Y. Qiu}
\affiliation{NIST center for Neutron Research, National Institute of Standard and Technology, Gaithersburg, Maryland 20899, USA}

\author{M. B. Maple}
\affiliation{Materials Science and Engineering Program, University of California, San Diego, La Jolla, California 92093, USA}
\affiliation{Department of Physics, University of California, San Diego, La Jolla, California 92093, USA}
\affiliation{Center for Advanced Nanoscience, University of California, San Diego, La Jolla, California 92093, USA}

\author{Dirk K. Morr}
\affiliation{ Department of Physics,
University of Illinois at Chicago, Chicago, Illinois 60607, USA}

\author{Pengcheng Dai}
\email{pdai@rice.edu}
\affiliation{ Department of Physics and Astronomy,
Rice University, Houston, Texas 77005, USA }

\begin{abstract}
The neutron spin resonance is a collective magnetic excitation that appears in copper oxide, iron pnictide, and
heavy fermion unconventional superconductors. Although the resonance is commonly associated with
a spin-exciton due to the $d$($s^{\pm}$)-wave symmetry of the superconducting order parameter, it has also been proposed to be a magnon-like excitation appearing in the superconducting state.
Here we use inelastic
neutron scattering to demonstrate that the resonance in the heavy fermion superconductor Ce$_{1-x}$Yb$_{x}$CoIn$_5$ with $x=0,0.05,0.3$ has a ring-like upward dispersion that is robust against Yb-doping. By comparing our experimental data with random phase approximation calculation using
the electronic structure and the momentum dependence of the $d_{x^2-y^2}$-wave superconducting gap determined from
scanning tunneling microscopy for CeCoIn$_5$, we conclude the robust upward dispersing resonance mode in Ce$_{1-x}$Yb$_{x}$CoIn$_5$ is inconsistent with the downward dispersion predicted within the spin-exciton scenario.
\end{abstract}

\maketitle


\section{introduction}
Understanding the origin of unconventional superconductivity in strongly correlated
electron materials continues to be at the forefront of modern condensed matter physics \cite{monthoux,scalapino,Keimer,Thompson,maple}.
In copper oxide \cite{Mignod,eschrig,Tranquada14}, iron pnictide \cite{Christianson,pcdai}, and heavy fermion \cite{Sato2001,CStock08} superconductors, the appearance of a neutron spin resonance below the
superconducting transition temperature $T_{\rm c}$ suggests that spin-fluctuation mediated pairing
is a common thread for different families of unconventional superconductors \cite{scalapino}.

The neutron spin resonance is a collective magnetic excitation coupled to superconductivity
with a temperature dependence similar to the
superconducting order parameter \cite{Mignod,eschrig}.  It is located near the antiferromagnetic (AF)
ordering wave vector ${\bf Q}_{\rm AF}$
of the undoped parent compound and its energy $E_{\rm r}$ at ${\bf Q}_{\rm AF}$ is
related to either $T_{\rm c}$ \cite{inosov11} or the superconducting energy gap $\Delta$ \cite{GYu09}. Although it is generally accepted that the resonance is a signature of unconventional superconductivity \cite{scalapino}, there is no consensus on its microscopic origin. A common interpretation of the resonance is that it is a spin-exciton,
arising from
particle-hole excitations involving momentum states near the Fermi surfaces
that possess opposite signs of the
$d$ (or $s^{\pm}$)-wave superconducting order parameter \cite{eschrig,Hirschfeld,CStock08}. Alternatively it has also been proposed to be a magnon-like excitation \cite{dkmorr,chubukov08}. At present, there is no consensus on its
microscopic origin \cite{scalapino,eschrig,Tranquada14,pcdai}. 

In hole-doped copper oxide superconductors, the magnetic excitations has an hourglass dispersion with a
downward dispersion at energies below $E_{\rm r}$ and an upward magnon-like  dispersion at energies above $E_{\rm r}$ \cite{Tranquada14}. The resonance, on the other hand, obtained by subtracting the normal state magnetic excitations from those in the superconducting state, displays predominantly
a downward dispersion \cite{Bourges,dai01,DReznik,Stock05}.
In the case of Ni-underdoped BaFe$_2$As$_2$ with coexisting AF order and superconductivity \cite{xylu13}, the resonance only
reveals an upward magnon-like dispersion \cite{MGKim13}. In both cases the resonance is well described by the spin-exciton scenario, the opposite dispersions being a result of $d_{x^2-y^2}$ or $s^\pm$ symmetry of the superconducting order parameter \cite{Eremin05,MGKim13}.

For the heavy fermion
superconductor CeCoIn$_5$ ($T_{\rm c}=2.3$ K) \cite{Thompson}, the resonance appears below $T_{\rm c}$ at $E_{\rm r}=0.60\pm 0.03$ meV and the
commensurate AF wave vector  ${\bf Q}_{\rm AF}=(1/2,1/2,1/2)$ in reciprocal space \cite{CStock08}.
Since CeCoIn$_5$ has a superconducting gap with $d_{x^2-y^2}$-wave symmetry as determined from scanning tunneling microscopy (STM) experiments \cite{MPAllan13,BBZhou}, the resonance is expected to show a downward dispersion similar to the cuprates within the spin-exciton picture \cite{Eremin08,Dyke14}.
Alternatively, the resonance, with its three-dimensional character \cite{CStock08}, could be a magnon-like excitation of $f$ electrons that becomes visible due to its reduced decay rate in the superconducting state \cite{dkmorr,chubukov08}. In this case, the resonance is not a signature of
$d_{x^2-y^2}$-wave superconductivity, but a measure of the hybridization between $f$ electrons and conduction electrons and its associated
pairing-sensitive Landau damping \cite{chubukov08}.

\begin{figure}[t]
\includegraphics[scale=.48]{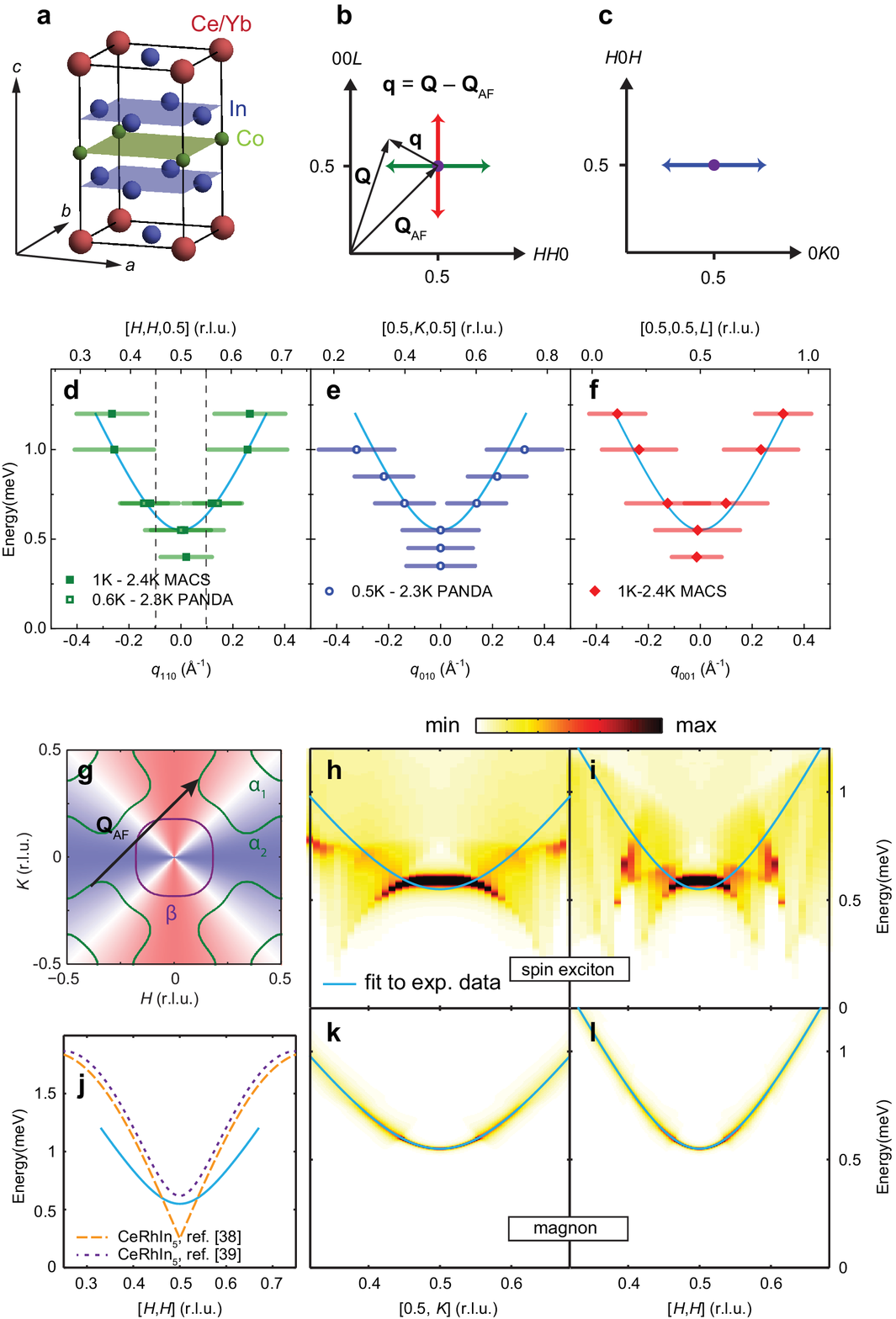}
\caption{
\textbf{Summary of neutron scattering results on Ce$_{0.95}$Yb$_{0.05}$CoIn$_{5}$.} (a) Crystal structure of Ce$_{1-x}$Yb$_{x}$CoIn$_{5}$.
(b) $[H,H,L]$ scattering plane, where
${\bf q}$ is measured from ${\bf Q}_{\rm AF}$ via ${\bf q}={\bf Q}-{\bf Q}_{\rm AF}$.
The red and green arrows represent scans along $[0.5,0.5,L]$ and $[H,H,0.5]$ centered at ${\bf Q}_{\rm AF}$,  respectively. (c) $[H,K,H]$ scattering plane. Here scans along $[0.5,K,0.5]$ centered at ${\bf Q}_{\rm AF}$
can be carried out as indicated by the blue arrow. (d) Dispersion of the resonance along $[H,H,0.5]$.
The axis above the figure is ${\bf Q}$ in r.l.u. whereas the axis at the bottom is ${\bf q}$ in \AA$^{-1}$.
An isotropic dispersion $E=\sqrt{\Delta^2+(c|{\bf q}|)^2}$ ($\Delta=0.55(1)$ meV, $c=3.2(1)$ meV$\cdot$\AA) is shown as a cyan solid line, where
$\Delta$ represents a spin gap and $c$ is the effective spin wave velocity.
The horizontal bars represent experimentally observed peak full-width-at-half-maximums (FWHM). The dashed vertical lines indicate the ordering wave vector of the so-called $Q$ phase at ${\bf Q}={\bf Q}_{\rm AF}\pm(\delta,\delta,0)$ with $\delta=0.05$ \cite{Kenzelmann10}. (e) and (f) are similar to (d) but are for dispersions along $[0.5,K,0.5]$ and $[0.5,0.5,L]$, respectively. (g) The Fermi surfaces of CeCoIn$_{5}$ where the blue and red shading represent the \textit{d}-wave symmetry of the superconductivity order parameter. The black arrow is ${\bf Q}_{\rm AF}$ that connects parts of Fermi surfaces with sign-reversed superconductivity order parameters. (h) Color-coded calculated intensity along the $[0.5,K]$ direction by considering the resonance mode to be a spin-exciton. (i) Calculated intensity for the spin-exciton along the $[H,H]$ direction. (j) Comparison of dispersions of the resonance in Ce$_{0.95}$Yb$_{0.05}$CoIn$_{5}$ (solid cyan line) and spin waves in CeRhIn$_{5}$ (dashed purple and orange lines) \cite{PDas,stock15}.
(k) Calculated intensity of the resonance along the $[0.5,K]$ direction assuming it is a magnon-like excitation. Dispersion of the magnon-like excitations are obtained from fits to experimental data and the intensity is affected by damping due to the particle-hole continuum. (l) Calculated intensity for the magnon-like excitation along the $[H,H]$ direction.
 }
\end{figure}

When La is substituted for Ce in Ce$_{1-x}$La$_{x}$CoIn$_5$ \cite{Petrovic02,Tanatar05},
superconductivity and the energy of the resonance are both rapidly suppressed while
$E_{\rm r}/k_{\rm B}T_{\rm c}$ remains approximately constant, where $k_{\rm B}$ is the Boltzmann constant \cite{Panarin,RaymondS11}.
At the same time, the energy width of the resonance
broadens with increasing La-doping \cite{Petrovic02,Tanatar05,Panarin,RaymondS11}.
When Yb is doped into CeCoIn$_5$ to form Ce$_{1-x}$Yb$_{x}$CoIn$_5$ superconductivity is suppressed much slower \cite{Lshu}. With increasing Yb,
de Haas-van Alphen and angle resolved photo emission studies find a change in the Fermi-surface topology between Yb nominal doping levels of $x=0.1$ and 0.2 \cite{Polyakov,Dudy}.
In addition, London penetration depth measurements
suggest that the superconducting gap changes from nodal to nodeless
around a similar Yb doping level \cite{Kim15}, arising possibly from composite electron pairing in a fully
gapped superconductor for $x>0.2$ \cite{Erten}.  If the resonance in CeCoIn$_5$ is a spin-exciton, it should
be dramatically affected by the Yb-doping induced changes in Fermi surface topology and superconducting gap.  On the other hand,
if the resonance is a magnon-like excitation, it should be much less sensitive to Yb-doping
across $x=0.2$ and display a upward dispersion similar to spin waves in
antiferromagnetically ordered nonsuperconducting CeRhIn$_5$ characteristic of a robust effective nearest-neighbor exchange coupling, regardless of its itinerant electron
or local moment origin \cite{eschrig,PDas,stock15}.

Here we use inelastic
neutron scattering to demonstrate that the resonance in the heavy fermion superconductor Ce$_{1-x}$Yb$_{x}$CoIn$_5$ with $x=0,0.05,0.3$ and $T_{\rm c}\approx 2.3, 2.25, 1.5$ K, respectively [Methods section and Supplementary Figure 1] \cite{Thompson,CStock08,Lshu} has a dominant ring-like upward dispersion that is robust against Yb-doping and the concomitant changes in electronic structure, a feature not present in the spin-exciton scenario. Moreover, a downward dispersion expected in the spin-exciton scenario is not observed. The robust upward dispersion of the resonance suggests it may have a magnon-like contribution \cite{chubukov08}. Specifically we find that the resonance in
Ce$_{0.95}$Yb$_{0.05}$CoIn$_5$ displays an upward dispersion along the $[H,H,0.5]$, $[0.5,K,0.5]$, and $[0.5,0.5,L]$ directions as shown in Fig. 1(d), 1(e), and 1(f), respectively. Upon increasing Yb-doping to $x=0.3$, the energy of the resonance at ${\bf Q}_{\rm AF}$ decreases corresponding to the reduction in $T_c$ [Supplementary Figure 2], but the overall dispersion and location of the mode in reciprocal space remain unchanged.
Upward dispersions similar to Ce$_{0.95}$Yb$_{0.05}$CoIn$_5$ are also found in undoped CeCoIn$_5$ and Ce$_{0.7}$Yb$_{0.3}$CoIn$_5$ [Supplementary Figures 3, 4 and 5].
Using the electronic structure and the momentum dependence of the $d_{x^2-y^2}$-wave superconducting gap determined from
STM for CeCoIn$_5$ [Fig. 1(g)] \cite{Dyke14}, we calculate the feedback of superconductivity on the magnetic excitations within the spin-exciton scenario [Supplementary Note 1, Supplementary Figures 6, 7 and 8].
The resulting wave vector dependence of the
spin-exciton along the $[0.5,K]$ and $[H,H]$ directions, which are shown in Fig. 1(h) and Fig. 1(i), respectively, are inconsistent with the experimentally determined upward dispersion (solid lines). 
Similar dispersive resonance in CeCoIn$_5$ and Ce$_{0.7}$Yb$_{0.3}$CoIn$_5$ [Figure 3, Supplementary Figures 3, 4 and Figure 5] are seen in spite of possible changes in the Fermi surface and superconducting
gap symmetry on moving from $x=0$ to 0.3 \cite{Polyakov,Dudy,Kim15}, also inconsistent with the expectation that a spin-exciton should depend sensitively on the Fermi surface.
We thus conclude that the upward dispersing resonance mode in
Ce$_{0.95}$Yb$_{0.05}$CoIn$_5$ cannot be interpreted as a spin-exciton arising from the feedback of unconventional $d$-wave superconductivity
\cite{CStock08,Eremin08,Dyke14}.
On the other hand, the similarity of the resonance's dispersion along the $[H,H,0.5]$ direction with
the spin-wave dispersion in AF ordered nonsuperconducting CeRhIn$_5$ along the same direction \cite{PDas,stock15} [Fig. 1(j)] suggests
the upward dispersing resonance may be magnon-like. In this scenario, the magnetic resonance arises since the opening of the superconducting gap leads to strong suppression of Landau damping for preexisting magnon-like excitations, as shown in Figs. 1(k) and 1(l) [Supplementary Note 2, Supplementary Figures 9, 10 and 11].
This is, therefore, the first experimental observation of a magnetic resonance in an unconventional superconductor that cannot be interpreted as a spin-exciton.

\begin{figure}[t]
\includegraphics[scale=.75]{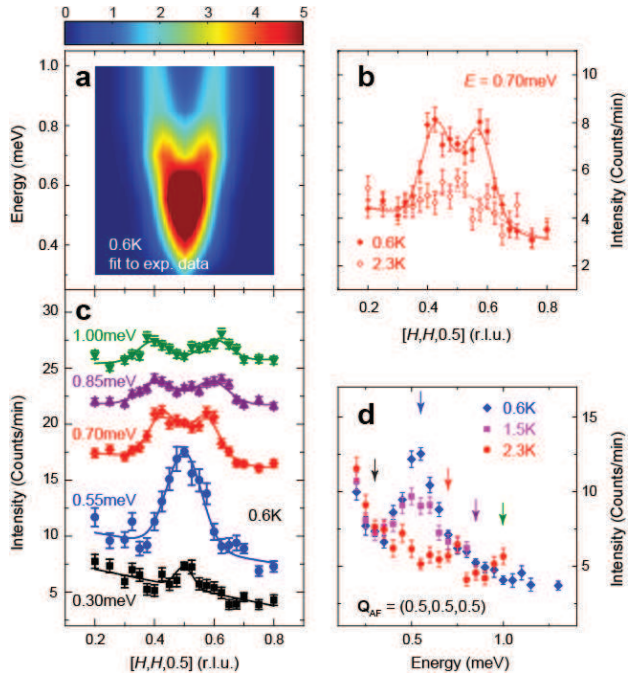}
\caption{
\textbf{Neutron scattering results on Ce$_{0.95}$Yb$_{0.05}$CoIn$_{5}$ in the $[H,H,L]$ scattering plane.} (a) Color-coded intensity of magnetic excitations along $[H,H,0.5]$ centered at ${\bf Q}_{\rm AF}$ at 0.6 K obtained from fits to data in (c). (b) Constant-energy scans along $[H,H,0.5]$ centered at ${\bf Q}_{\rm AF}$ with $E = 0.7$ meV. The solid symbols are data well below $T_{\rm c}$ (0.6 K) where two peaks can be resolved whereas open symbols are obtained above $T_{\rm c}$ (2.3 K) showing a single peak centered at ${\bf Q}_{\rm AF}$. The solid line is a fit to the data at 0.6 K with two Gaussian functions whereas the dashed line is a fit to a single Gaussian function for the data at 2.3 K. Data at the two temperatures are fit simultaneously to have the same linear background. (c) Constant-energy scans along $[H,H,0.5]$ at 0.6 K. For clarity, scans with $E = 0.55, 0.75,0.75$ and $1$ meV are respectively shifted upwards by $5, 13, 18$ and $22$. The solid lines are fits to either one or two Gaussian functions with a linear background. (d) Constant-${\bf Q}$ scans at ${\bf Q}_{\rm AF}$. The arrows represent energies at which constant-energy scans are shown in (c). All vertical error bars in the figure represent statistical error of 1 standard deviation.
 }
\end{figure}

\section{Results}
\subsection{Dispersion of the resonance in Ce$_{0.95}$Yb$_{0.05}$CoIn$_5$ along $[H,H,0.5]$ and $[0.5,0.5,L]$}
Using a tetragonal unit cell with $a=b=4.60$ \AA, and $c=7.51$ \AA\ for Ce$_{0.95}$Yb$_{0.05}$CoIn$_5$ [Fig. 1(a)],
we define the momentum transfer ${\bf Q}$ in three-dimensional reciprocal space in \AA$^{-1}$
as ${\bf Q}=H{\bf a}^\ast+K{\bf b}^\ast+L{\bf c}^\ast$, where $H$, $K$, and $L$ are Miller indices and
${\bf a}^\ast=\hat{{\bf a}}2\pi/a$, ${\bf b}^\ast=\hat{{\bf b}}2\pi/b$,
${\bf c}^\ast=\hat{{\bf c}}2\pi/c$.
The experiments are carried out using the $[H,H,L]$ and $[H,K,H]$ scattering planes
to study the dispersions of the resonance along $[H,H,0.5]$, $[0.5,K,0.5]$, and $[0.5,0.5,L]$ [Figs. 1(b) and 1(c)]. Figure 2(a) shows the color-coded plot
of the spin excitations at 0.6 K obtained from fits to the raw data
at energies $E=0.3, 0.55, 0.7, 0.85,$ and 1 meV along $[H,H,0.5]$ for Ce$_{0.95}$Yb$_{0.05}$CoIn$_{5}$ [Fig. 2(c)].
While the data show a weak commensurate peak at $E=0.3$ meV, we see a
clear commensurate resonance at $E_{\rm r}\approx 0.55$ meV and upward dispersing incommensurate peaks
for energies $E=0.7, 0.85, 1$ meV.  Figure 2(b) shows constant-energy scans at $E=0.7$ meV
below and above $T_{\rm c}$.  At $T=2.3$ K, we see a broad peak centered
at the commensurate AF wave vector  ${\bf Q}_{\rm AF}$.  Upon cooling to below $T_{\rm c}$ at
$T=0.6$ K, the commensurate peak becomes two incommensurate peaks which disperse outward
with increasing energy [Fig. 2(c)].  Figure 2(d) shows constant-${\bf Q}$
scans at ${\bf Q}_{\rm AF}$ for temperatures $T=0.6, 1.5,$ and 2.3 K.
Similar to previous work on pure CeCoIn$_5$ \cite{CStock08}, the data reveal a clear resonance
at $E_{\rm r}\approx 0.55$ meV below $T_{\rm c}$, and no peak in the normal state above $T_{\rm c}$.

\begin{figure}[t]
\includegraphics[scale=.5]{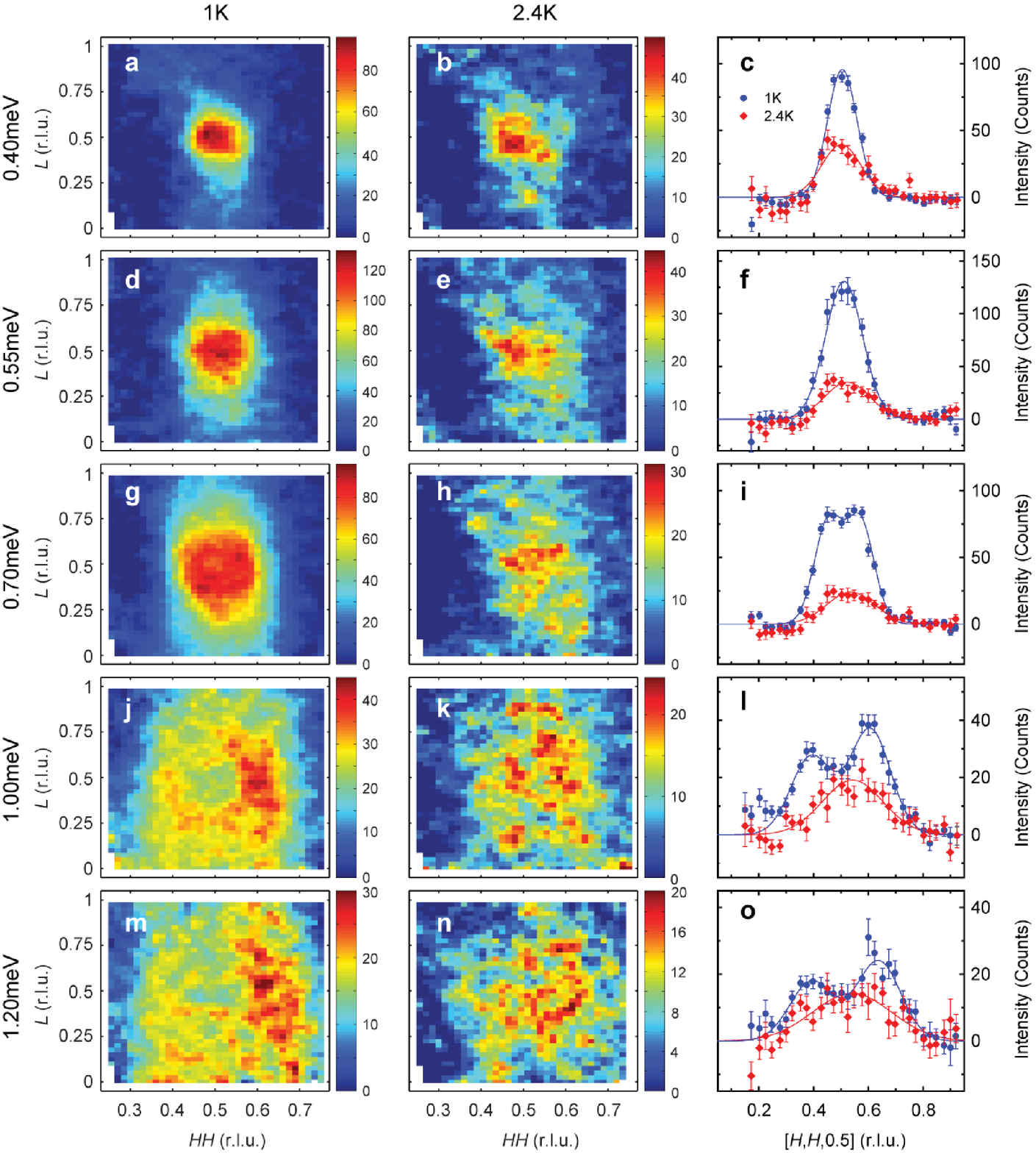}
\caption{
\textbf{Constant-energy maps of scattering intensity in the $[H,H,L]$ scattering plane for Ce$_{0.95}$Yb$_{0.05}$CoIn$_5$.}
Constant-energy map at $E=0.40$ meV at (a) 1 K and (b) 2.4 K.
A $|{\bf Q}|$-dependent background has been subtracted. (c) Cuts obtained from (a) and (b) by
binning data with $ 0.45 \leq L \leq 0.55$; solid lines are fits to the data using either a single or two Gaussian functions. Since a
background has already been subtracted in maps in (a) and (b), no background is assumed in the fits.
Similarly, (d), (e) and (f) are for $E = 0.55$ meV, (g), (h) and (i) are for $E=0.70$ meV, (j), (k) and (l) are for $E=1.00$ meV and (m), (n) and (o) are for $E = 1.20$ meV. All vertical error bars in the figure represent statistical error of 1 standard deviation.
 }
\end{figure}

To further illustrate the dispersive nature of the resonance, we show in Figure 3 maps in the
$[H,H,L]$ scattering plane of the spin excitations at different energies above and below $T_{c}$ obtained on MACS for
Ce$_{0.95}$Yb$_{0.05}$CoIn$_{5}$.
In the probed reciprocal space, we see clear spin excitations around ${\bf Q}_{\rm AF}$ which disperse
outward with increasing energy.  At an energy ($E=0.4$ meV) below the resonance, spin excitations
are commensurate
below [Fig. 3(a)] and above [Fig. 3(b)] $T_{\rm c}$.  The constant-energy cuts of the data
along the $[H,H,0.5]$ direction confirm this conclusion [Fig. 3(c)].
Figures 3(d), 3(e), and 3(f) show similar scans at $E=0.55$ meV and indicate that the scattering become broader in reciprocal space. Upon moving to $E=0.7$ meV [Figs. 3(g), 3(h), 3(i)],
1.0 meV [Figs. 3(j), 3(k), 3(l)], and 1.2 meV [Figs. 3(m), 3(n), 3(o)], we see clear ring-like
scattering dispersing away from ${\bf Q}_{\rm AF}$ with increasing energy in the superconducting state.
The normal state scattering is commensurate at all energies, and this is most clearly seen in the
constant-energy cuts along the $[H,H,0.5]$ direction. Based on the difference of data at 2.1 K and 1 K in Figs. 3, one can compose the
dispersions of the resonance along the $[H,H,0.5]$ [Fig. 1(d)] and $[0.5,0.5,L]$ [Fig. 1(f)] directions.
By plotting the dispersion in \AA$^{-1}$ away from ${\bf Q}_{\rm AF}$ [${\bf q}$ as defined in Fig. 1(b)], we see
that the resonance disperses almost isotropically along these two directions.

\begin{figure}[t]
\includegraphics[scale=0.55]{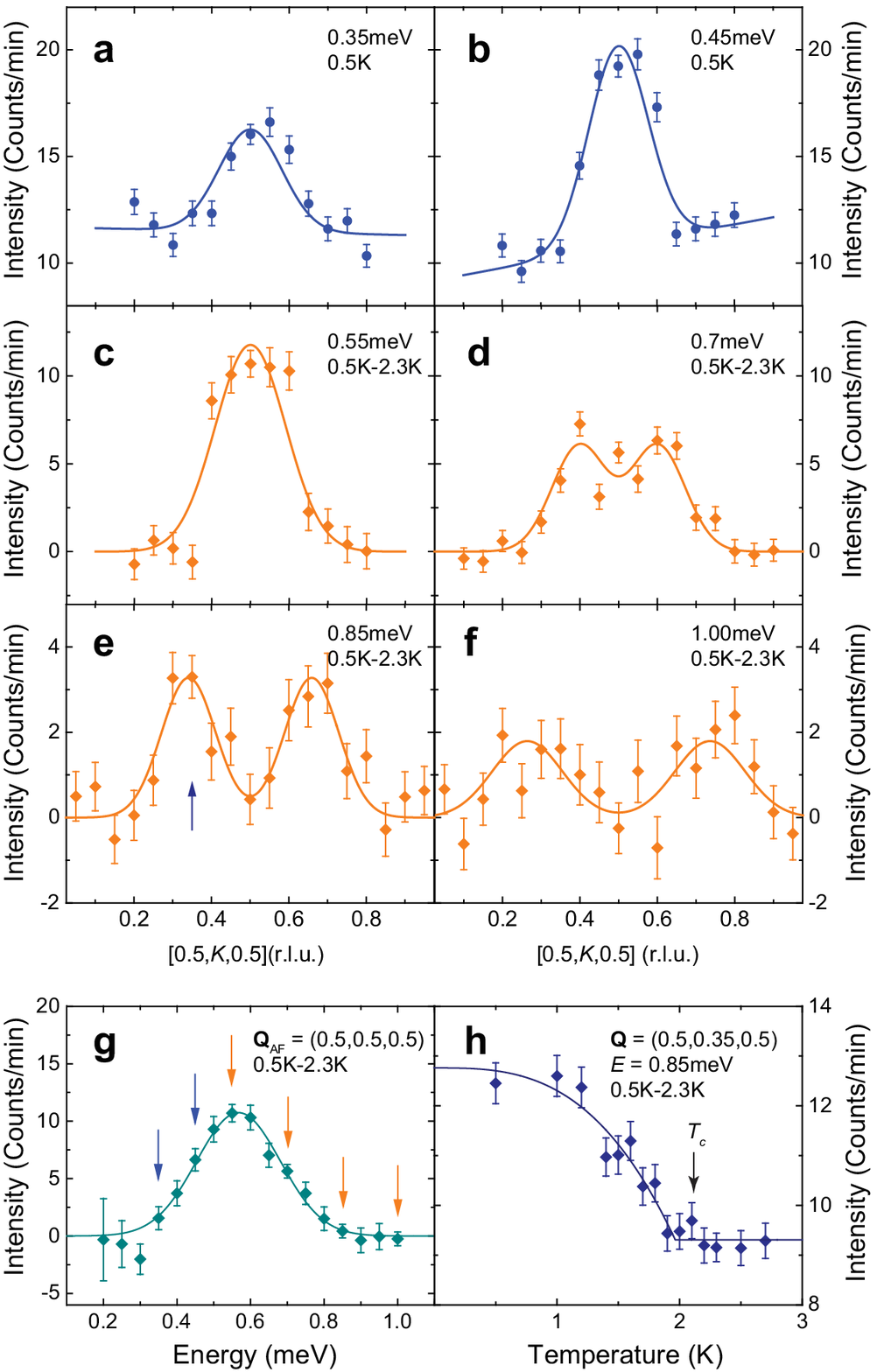}
\caption{
\textbf{Neutron scattering results on Ce$_{0.95}$Yb$_{0.05}$CoIn$_{5}$ in the $[H,K,H]$ scattering plane.} (a) Constant-energy scan along $[0.5,K,0.5]$ centered at ${\bf Q}_{\rm AF}$ at 0.5 K for $E=0.35$ meV. The solid line is a fit to a single Gaussian with a linear background. (b) Similar to (a) but for $E=0.45$ meV. (c) Constant-energy scan along $[0.5,K,0.5]$ centered at ${\bf Q}_{\rm AF}$, obtained by subtracting data at 2.3 K from data at 0.5 K for $E=0.55$ meV. The solid line is a fit to a Gaussian function with zero background. (d) Similar to (c), but for $E=0.7$ meV and the solid line is a fit to two Gaussian functions. (e) Similar to (d), but for $E=0.85$ meV. The arrow points to ${\bf Q}=(0.5,0.35,0.5)$, where measurement of the temperature dependence was carried out, shown in (h). (f) Similar to (d) and (e), but for $E=1.00$ meV. (g) Constant-${\bf Q}$ scan at ${\bf Q}_{\rm AF}$ obtained by subtracting the 2.3 K data from the 0.5 K data. The solid line is a Gaussian function centered at $E=0.57(1)$ meV with zero background. Arrows represent energies at which constant-energy scans are shown in (a)-(f). (h) Temperature dependence of scattering intensity at ${\bf Q}=(0.5,0.35,0.5)$ for $E=0.85$ meV. The solid line is a fit to $d$-wave superconductivity order parameter with constant background. The superconducting critical temperature $T_{\rm c}$ obtained from the fit is 2.0(1) K. All vertical error bars in the figure represent statistical error of 1 standard deviation.
}
\end{figure}

\subsection{Dispersion of the resonance in Ce$_{0.95}$Yb$_{0.05}$CoIn$_5$ along $[0.5,K,0.5]$}
In cuprate superconductors such as YBa$_2$Cu$_3$O$_{6.5}$ \cite{Stock05}, YBa$_2$Cu$_3$O$_{6.6}$ \cite{smhayden}, and La$_{1.875}$Ba$_{0.125}$CuO$_4$ \cite{Tranquada04},
spin excitations above the resonance form a ring-like upward dispersion in the $ab$ plane slightly softened from the spin waves in their AF ordered parent compounds \cite{Tranquada14}.  To conclusively determine if the resonance dispersion is also ring-like in the $ab$ plane in Ce$_{0.95}$Yb$_{0.05}$CoIn$_5$, we aligned the single crystals in the $[H,0,H]\times [0,K,0]$ ($[H,K,H]$) scattering plane to measure the dispersion of the resonance along $[0.5,K,0.5]$ centered at ${\bf Q}_{\rm AF}$.
Figure 4(a)-4(f) summarizes the constant-energy scans at $E=0.35, 0.45, 0.55, 0.7, 0.85, 1.0$ meV along
$[0.5,K,0.5]$.  While the scattering is clearly
commensurate at $E=0.35, 0.45$ meV below the resonance at $E_{\rm r}\approx 0.55$ meV [Fig. 4(a) and 4(b)],
it becomes incommensurate above the resonance at $E= 0.7, 0.85, 1.0$ meV with
an upward dispersion as a function of increasing energy [Figs. 4(d), 4(e), and 4(f)].
Figure 1(e) summarizes the dispersion of the resonance in \AA$^{-1}$ away
from ${\bf Q}_{AF}$ along $[0.5,K,0.5]$.
Figure 4(g) shows
the difference of the constant-${\bf Q}$ scans below and above $T_{\rm c}$ at ${\bf Q}_{\rm AF}$, again
revealing a strong peak at the resonance energy of $E_{\rm r}\approx 0.55$ meV similar to Fig. 2(d).
Finally,  Fig. 4(h) shows temperature dependence of the scattering
at an incommensurate wave vector $(0.5,0.35,0.5)$ and $E=0.85$ meV, which
reveals a clear superconducting order-parameter-like increase below $T_{\rm c}$ and indicates
that the incommensurate part of the resonance is also coupled to superconductivity.

\begin{figure}[t]
\includegraphics[scale=0.7]{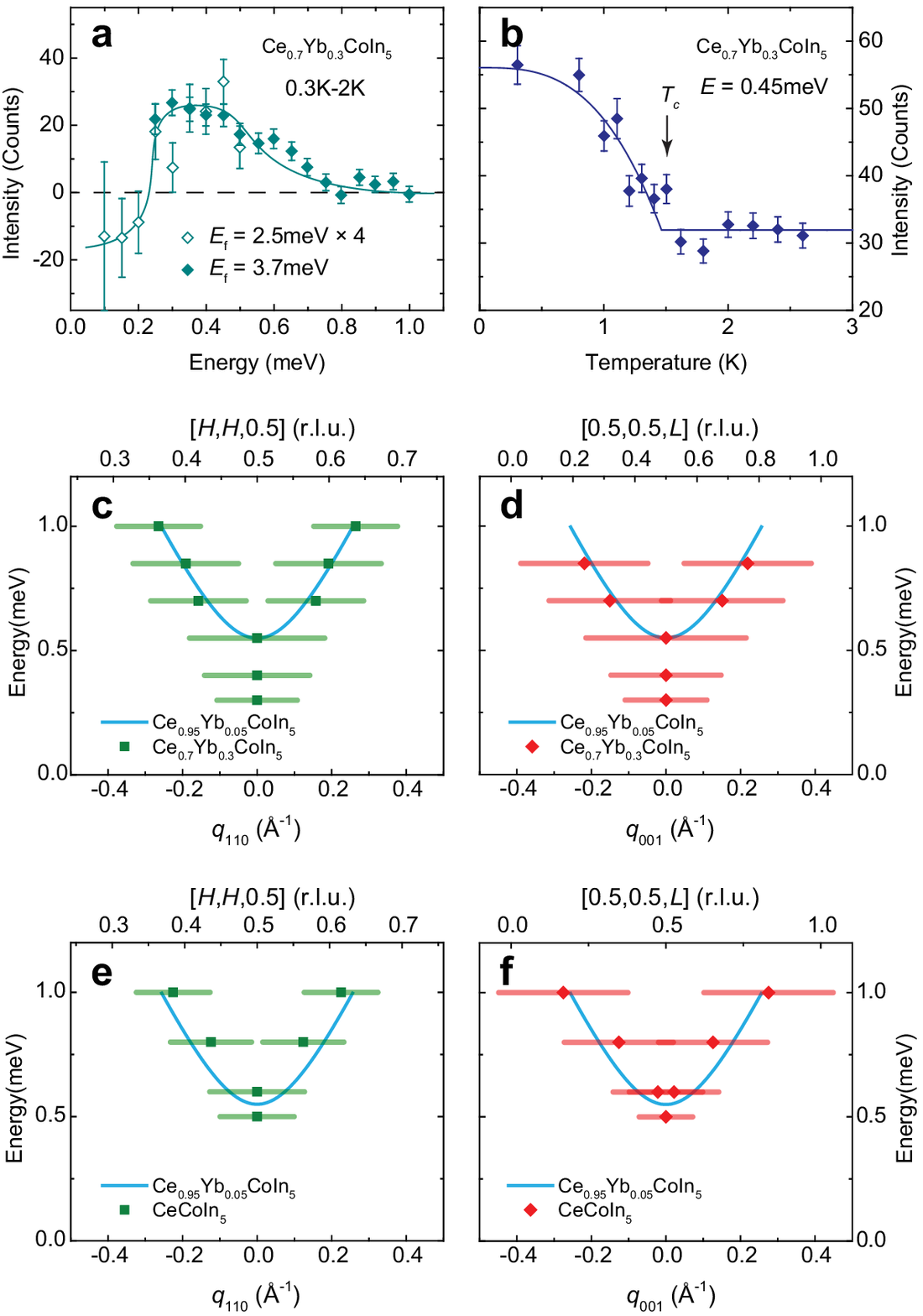}
\caption{
\textbf{Summary of neutron scattering results on Ce$_{0.7}$Yb$_{0.3}$CoIn$_{5}$ and CeCoIn$_5$.} (a) Difference of constant-${\bf Q}$ scans at ${\bf Q}_{\rm AF}$=(0.5,0.5,0.5) for 0.3 K and 2 K displaying a resonance mode at $E_{\rm r} \approx 0.4$ meV for Ce$_{0.7}$Yb$_{0.3}$CoIn$_{5}$. Filled symbols are obtained with fixed scattered neutron energy $E_{\rm f} = 3.7$ meV and open symbols are for $E_{\rm f} = 2.5$ meV scaled up by 4 times. All of the data in the rest of figure are obtained with $E_{\rm f} = 3.7$ meV. The solid line is a guide to the eye. (b) Temperature dependence of the resonance mode in Ce$_{0.7}$Yb$_{0.3}$CoIn$_{5}$
for $E = 0.45$ meV and ${\bf Q}_{\rm AF}$=(0.5,0.5,0.5), the solid line is a fit to $d$-wave superconducting gap, with $T_{\rm c} = 1.5(1)$ K.
Dispersion of the resonance along (c) $[H,H,0.5]$ and (d) $[0.5,0.5,L]$
for Ce$_{0.7}$Yb$_{0.3}$CoIn$_{5}$.
Dispersions of the resonance for CeCoIn$_5$ along $[H,H,0.5]$ and $[0.5,0.5,L]$  are showin in (e) and (f), respectively. The solid cyan lines in (c)-(f) are dispersions of the resonance obtained Ce$_{0.95}$Yb$_{0.05}$CoIn$_{5}$. The horizontal bars represent experimentally observed peak full-width-at-half-maximums (FWHM). All vertical error bars in the figure represent statistical error of 1 standard deviation.
}
\end{figure}

\subsection{Dispersion of the resonance for CeCoIn$_5$ and Ce$_{0.7}$Yb$_{0.3}$CoIn$_5$}
To determine how Yb-doping, and in particular the possible changes in the Fermi surface topology
and superconducting gap structure between Yb-doping of $x=0.1$ and $x=0.2$
affect the behavior of the resonance \cite{Polyakov,Dudy,Kim15}, we carried out additional
inelastic neutron scattering experiments on CeCoIn$_5$ and Ce$_{0.7}$Yb$_{0.3}$CoIn$_{5}$ at MACS.  Figure 5(a) shows
temperature differences of constant-{\bf Q} scans at ${\bf Q}_{\rm AF}$ below and above $T_{\rm c}$ in Ce$_{0.7}$Yb$_{0.3}$CoIn$_5$, which reveals a clear resonance at $E_{\rm r} \approx 0.4$ meV.
Figure 5(b) plots the temperature dependence of the resonance, displaying a superconducting order-parameter-like increase
in intensity below $T_{\rm c}$.  From wave vector scans along the $[H,H,0.5]$ and $[0.5,0.5,L]$ directions
at different energies below and above $T_{\rm c}$ for Ce$_{0.7}$Yb$_{0.3}$CoIn$_{5}$ [Supplementary Figure 5], we can establish the dispersions of
the resonance along these two directions as shown in Figures 5(c) and 5(d), respectively.
Similarly, Figures 5(e) and 5(f) compare dispersions of the resonance for
CeCoIn$_5$ [Supplementary Figure 4] and Ce$_{0.95}$Yb$_{0.05}$CoIn$_5$ along the $[H,H,0.5]$ and $[0.5,0.5,L]$ directions, respectively.
From Figs. 5(c)-5(f), we see that the dispersions of the resonance are essentially Yb-doping independent.
However, the bottom of dispersive resonance at ${\bf Q}_{\rm AF}$ moves down in energy
with increasing Yb-doping and $E_{\rm r}$ is proportional to $k_{\rm B}T_{\rm c}$, similar to La-doped
CeCoIn$_5$ \cite{Panarin,RaymondS11}.

\section{Discussion}
From the dispersions of the resonance
along $[H,H,0.5]$ [Fig. 1(d)], $[0.5,K,0.5]$ [Fig. 1(e)],
and $[0.5,0.5,L]$ [Fig. 1(f)] for Ce$_{0.95}$Yb$_{0.05}$CoIn$_{5}$, we see that the mode disperses isotropically
in reciprocal space away from ${\bf Q}_{\rm AF}$, 
which is inconsistent with the resonance being a spin-exciton [see Figs.1(h) and 1(i)], but resembles a magnon-like excitation with a dispersion resembling spin waves in CeRhIn$_5$[Fig. 1(j), Supplementary Note 3 and Supplementary Figure 12] that becomes undamped in the superconducting state \cite{dkmorr,chubukov08}. However, 
the fact that CeCoIn$_5$ is a multiband system complicates the identification of the resonance's origin. While we assumed here that the main contribution to the resonance arises from the quasi-localized $f$-levels identified via quasi-particle interference (QPI) spectroscopy in STM experiment \cite{MPAllan13,Dyke14}, it is of course possible that there exist further electronic bands that become superconducting and contribute to the resonance (either directly or through a renormalization of the magnetic interaction) but were not detected via QPI spectroscopy. Clearly, further studies are necessary to investigate this possibility.

Moreover, in a recent work on pure CeCoIn$_5$, it was suggested that
the resonance in the energy range of 0.4-0.7 meV is incommensurate
along the $[H,H,0.5]$ direction with wave-vector ${\bf Q}_{\rm AF}\pm (\delta,\delta,0)$ where $\delta=0.042(2)$ r.l.u. \cite{raymond15}.
Since the incommensurate
wave vectors of the resonance appear to be close to the in-plane magnetic field induced incommensurate static
magnetic order at ${\bf Q}_{\rm AF}\pm (\delta,\delta,0)$
with $\delta=0.05$ (the so-called $Q$ phase) [see the vertical dashed lines
in Fig. 1(d)] \cite{Kenzelmann08,Kenzelmann10,Gerber14}, and since it was suggested that the fluctuating moment of the resonance is entirely polarized along the $c$-axis similar to the ordered moment of the
$Q$ phase \cite{CStock08,raymond15}, the resonance
has been described as a dynamical precursor of the $Q$ phase \cite{Michal}. 
Experimentally, we did not observe incommensurate excitations at $E = 0.5$ meV, nevertheless our data suggest a smaller splitting than in previous work if the excitations at $E=0.5$ are incommensurate [Supplementary Note 4 and Supplementary Figure 13]. Furthermore, the $Q$ phase precursor interpretation of the resonance is also inconsistent with the observed ring-like dispersion at $E>0.7$ meV. It is possible that there are more than one contribution to the resonance in CeCoIn$_5$ given its electronic complexity. In the present work, we identify the upward dispersing magnon-like contribution being dominant, but do not rule out finer features at lower energies with $E<0.6$ meV which can only be resolved with better resolution.  Our data and previous work on CeCoIn$_5$ \cite{raymond15}
are consistent with each other, both showing no signature of a downward dispersion. 

Further insight into the nature of the resonance in CeCoIn$_5$ can be gained by considering its behavior in an applied magnetic field. Previous neutron scattering experiments by Stock {\it et al.} \cite{Stock2012} observed that the resonance in the superconducting state of CeCoIn$_5$ splits into two modes if a magnetic field is applied along the $[1,\bar{1},0]$ direction. This splitting into two modes by an in-plane field is rather puzzling, since for a system with a Heisenberg spin symmetry, a splitting into three modes is expected. Moreover, if the resonance in CeCoIn$_5$ was entirely polarized along the $c$-axis \cite{CStock08,raymond15},
application of an in-plane magnetic field should not split the resonance into the doublet observed
experimentally \cite{Stock2012,Raymond2012}.
However, this observation can be explained if the system possess a magnetic anisotropy with a magnetic easy-plane [indicated by the green ellipse in
Fig.~6(a)] that is perpendicular to the direction of the applied magnetic field [red arrow in Fig.~6(a)].  Since the magnetic field applied by Stock {\it et al.} \cite{Stock2012} lies in the $[1,{\bar 1},0]$ direction, this implies that the easy plane is spanned by the unit vectors in the $[0,0,1]$ and $[1,1,0]$ directions. This leads us to suggest that the resonance
in CeCoIn$_5$ should also have a component along the $[1,1,0]$ direction in addition to the $c$-axis component similar to
the resonance in electron-doped iron pnictides \cite{Steffens,HQLuo13}.  Such in-plane spin excitation anisotropy can occur due to the presence of
spin-orbit coupling, and does not break the four-fold rotational symmetry of the underlying lattice \cite{HQLuo13}. Present experimental results do not rule the presence of such a mode, although it is also challenging to experimentally confirm its presence [Supplementary Note 5, Supplementary Figures 14 and 15]. 

\begin{figure}
\begin{center}
 \includegraphics[height=5cm]{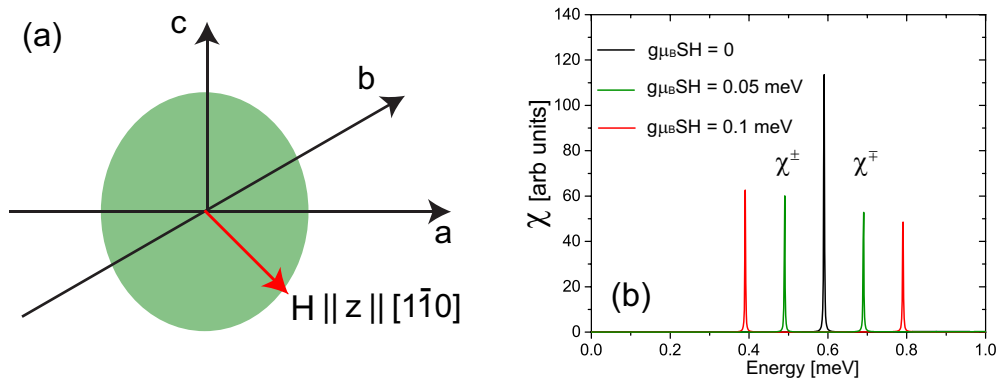}
 \caption{\textbf{Effect of applied magnetic field on the resonance mode.} (a) Orientation of the magnetic field {\bf H} and of the magnetic easy plane in the crystal lattice. The magnetic field is perpendicular to the magnetic easy plane. (b) Evolution of the resonance with increasing magnetic field magnetic.}
 \label{fig:Im_chi_magfield}
  \end{center}
 \end{figure}

To quantitatively understand the effect of a magnetic field on spin excitations, we consider the Hamiltonian (see Supplementary Eq. 1 in Ref.~\cite{Dyke14})
\begin{align}
H&=\sum_{{\bf r,r'}} I_{{\bf r,r'}} {\bf S}_{\bf r} \cdot {\bf S}_{\bf r'} + A \sum_{\bf r} \left( S_{\bf r}^z \right)^2  - g \mu_{\rm B} H \sum_{\bf r} S_{\bf r}^z
\end{align}
with the three terms representing the magnetic interactions between the $f$-electron moments, the magnetic anisotropy of the system and the interaction with the external magnetic field, respectively. Here, we define the direction of the magnetic field along the $[1,{\bar 1},0]$ direction as the $z$-axis in spin space. We assume $A>0$, such that the system possesses a hard magnetic axis along $[1,{\bar 1},0]$ and an easy plane [green ellipse in Fig.~\ref{fig:Im_chi_magfield}(a)] perpendicular to it. This Hamiltonian implies that the effective interaction for the longitudinal, non-spin-flip scattering mode (parallel to the applied field) is given by $I_{zz}({\bf q}) = I_{\bf q} + A$, while the interaction for the transverse mode is given by $I_{\pm}({\bf q}) = I_{\bf q}$
with $I_{\bf q}$ being the Fourier transform of $I_{{\bf r,r'}}$  in Eq. (1).
In the vicinity of the AF wave-vector ${\bf Q}_{\rm AF}$, where  $I_{{\bf Q}_{\rm AF}}<0$, we thus obtain
$|I_{zz}({\bf Q}_{\rm AF})|< |I_{\pm}({\bf Q}_{\rm AF})|$
since $A>0$ for an easy-plane perpendicular to the $[1,{\bar 1},0]$ direction. This implies that the effective interaction at ${\bf Q}_{\rm AF}$ for the longitudinal, non-spin-flip scattering mode (parallel to the applied field) is smaller than in the two transverse, spin-flip scattering modes which lie in the easy plane. As a result, the longitudinal mode will be located at energies higher than the transverse modes. In particular, for sufficiently large $A$ the longitudinal mode can be located above the onset energy, $\omega_c({\bf Q}_{\rm AF})$, for the particle-hole continuum in the superconducting state, and thus would not emerge as a resonance peak. Hence, only the two transverse modes within the easy plane contribute to the resonance peak. The application of a magnetic field perpendicular to the easy-plane of the system then splits the two transverse modes of the resonance peak in energy (while not affecting the longitudinal mode), with the energy splitting increasing linearly with the magnetic field, as shown in Fig.~6(b), thus explaining the experimental observation in Ref.~\cite{Stock2012}.

If spin excitations in CeCoIn$_5$ are only polarized along the $c$-axis with
the existence of an easy axis rather than an easy plane \cite{CStock08,raymond15}, application of a magnetic field along the direction perpendicular
to the easy axis along the $[1,{\bar 1},0]$ direction, the transverse mode along the easy axis shifts down with increasing field, but does not split.  Similarly,
when a field is applied along the easy axis direction ($c$-axis field), the two transverse modes are located at higher energies, while the longitudinal mode, which is located at lower energies, does not split in the magnetic field.  The presence of a longitudinal spin excitations along the $[1,1,0]$ direction is also consistent the magnetic field effect work of Ref. \cite{Raymond2012}, where the resonance is believed to be a
composite excitation which contains three excitation channels involving
both transverse and longitudinal modes.

While unconventional superconductivity in copper oxide, iron pnictide, and heavy fermion superconductors appears with the suppression of the static AF order in their parent compounds, dispersive magnon-like excitations persist in the doped superconductors \cite{Tranquada14,pcdai,Stockert}.
Our discovery that the resonance itself in Ce$_{1-x}$Yb$_{x}$CoIn$_5$ shows robust ring-like upwards dispersion suggests instead of a spin-exciton in a $d$-wave superconductor \cite{scalapino,eschrig},
the resonance may be a magnon-like excitation revealed in the superconducting state \cite{chubukov08}. Since the presence of a propagating spin resonance is characteristic of a nearby AF state, we propose that the magnon-like resonance mode in
Ce$_{1-x}$Yb$_{x}$CoIn$_5$ is the strong-coupling analogue of a weak coupling spin-exciton. This would imply that the nature of the magnetic resonance -- spin-exciton versus magnon-like excitations -- represents a new criterion to distinguish between more weakly and more strongly coupled unconventional superconductors.

{\bf Methods}

{\bf Sample preparation}
Single crystals of Ce$_{1-x}$Yb$_{x}$CoIn$_{5}$ ($x$ = 0, 0.05, 0.3) were prepared by indium self-flux method. Details of sample preparation and characterizations have been previously reported, lattice parameters for Ce$_{1-x}$Yb$_{x}$CoIn$_{5}$ remain similar to pure CeCoIn$_{5}$ for all reported doping levels \cite{Lshu}. We use the nominal doping throughout the paper to be consistent with earlier work \cite{Lshu}, while the actual doping is $\sim$1/3 of the nominal doping \cite{SJang}. Supplementary Fig. 1(a) shows the out-of-phase AC magnetic susceptibility (15.9 Hz) measured on Ce$_{1-x}$Yb$_x$CoIn$_5$ samples with $x$ = 0.05 and 0.3 from the same growth batches used for neutron scattering experiments. Bulk superconductivity appear at $T_c$ = 2.25 K and $T_c = 1.5$ K respectively, whereas $T_c = 2.3$ K in pure CeCoIn5\cite{Lshu}.

Hundreds of Ce$_{1-x}$Yb$_x$CoIn$_5$ single crystals with total masses of ~0.8 g, ~2.5 g and ~1.4 g respectively for $x$ =
0, 0.05 and 0.3 were co-aligned on several aluminum plates using CYTOP as hydrogen-free glue [Supplementary Fig. 1(b)]. The
plates are then mounted in either the $[H,H,0]\times[0,0,L]$ ($[H,H,L]$) [Supplementary Fig. 1(c)] or the $[H,0,H]\times[0,K,0]$ ($[H,K,H]$) scattering plane [Supplementary Fig. 1(d)]. The total thickness of samples on co-aligned plates is 1-2 mm, minimizing neutron absorption due to indium. Absorption becomes most significant when the incident or the scattered neutron beam becomes perpendicular to $[0,0,1]$, which does not occur for reciprocal space regions shown in this work.

{\bf Experiment details and analysis}
Neutron scattering experiments were carried out on the PANDA cold three-axes spectrometer at Heinz Maier-Leibnitz Zentrum and the Multi-Axis Crystal Spectrometer (MACS) at the NIST Center for Neutron Research. The experiments on PANDA used a Be filter 180 mm in length after the sample which is highly effective in removing contamination from higher order neutrons, both the analyzer and the monochromator are doubly focused to maximize neutron flux at the sample. Vertical focusing of the analyzer is fixed whereas horizontal focusing is variable. Both the horizontal and vertical focusing of the monochromator are variable. The variable focusings are adjusted depending on the neutron wavelength based on empirically optimized values. The PANDA experiment in the $[H,H,L]$ scattering plane used fixed $k_{\rm f}=1.3$ \AA$^{-1}$ ($E_{\rm f}\approx 3.5$ meV) and the experiment in the $[H,K,H]$ scattering plane used fixed $k_{\rm f}=1.57$ \AA$^{-1}$ ($E_{\rm f}\approx 5.1$ meV). The MACS experiments in the $[H,H,L]$ scattering plane used Be filters both before and after the sample with fixed $E_{\rm f} = 3.7$ meV. MACS consists of 20 spectroscopic detectors each separated by 8$^\circ$. By rotating the sample and shifting all of the detectors to bridge the 8$^\circ$ gaps, a map in terms of sample rotation angle and scattering angle at a fixed energy transfer can be efficiently constructed. A significant portion of the reciprocal space in the
scattering plane can be covered, which further allows cuts along the high symmetry directions. 90$^\prime$ collimators are used between the sample and each individual analyzers. The analyzers are vertically focused while the monochromator is doubly focused.

For the neutron scattering results on PANDA, a linear background is assumed for all measured constant-energy scans while no background is used for scans obtained by subtraction data above $T_{\rm c}$ from those obtained below $T_{\rm c}$. The constant energy scans are then simply fit to either one or two Gaussian peaks.
For the neutron scattering results obtained on MACS, maps of large portions of the scattering plane for several energies transfers were collected both below and above $T_{\rm c}$. A $|{\bf Q}|$-dependent background is obtained by masking the signal near $(0.5,0.5,0.5)$ and is then fit to a polynomial. The signal with $|{\bf Q}|<0.5$ \AA$^{-1}$ is masked throughout the analysis. The fit background is then subtracted from the map and the data is folded into the first quadrant of the scattering plane to improve statistics. The results for Ce$_{0.95}$Yb$_{0.05}$CoIn$_{5}$ are shown in Figure 3 and Supplementary Figure 3. Cuts along $[H,H,0.5]$ are obtained by binning data with $0.45\leq L \leq 0.55$  and fit with a single or two Gaussian peaks. Cuts along $[0.5,0.5,L]$ are obtained by binning data with $0.45\leq H \leq 0.55$ and fit by a sum of Lorentzian peaks, accounting for the Ce$^{3+}$ magnetic form factor $f({\bf Q})$ and the polarization factor assuming excitations are dominantly polarized along the $c$-axis similar to previous work \cite{CStock08}. The possible presence of excitations polarized along the $[1,1,0]$ direction is discussed in Supplementary Note 5. The function used to fit scans along $[0.5,0.5,L]$ can be written as
\begin{equation}
I({\bf Q})\propto f({\bf Q})^2 (1-(\hat{\bf Q} \cdot \hat{\bf c})^2) \sum_{n=-\infty}^{\infty}{F(n+L)}
\end{equation}
where $F(L)$ is either a single Lorentizan peak centered at $L=0.5$ or two Lorentzian peaks equally displaced from $L=0.5$. The peaks along $[0.5,0.5,L]$ are significantly broader compared to those along $[H,H,0.5]$, and remains non-zero even for $L=0$ [Supplementary Figure 3]. This contrasts with similar scans along $[H,H,0.5]$ in Fig. 3 where the intensity drops to zero away from ${\bf Q}_{\rm AF}$. MACS data of CeCoIn$_{5}$ and Ce$_{0.7}$Yb$_{0.3}$CoIn$_{5}$ with the corresponding maps and cuts are shown Supplementary Figures 4 and 5. Similar to Ce$_{0.95}$Yb$_{0.05}$CoIn$_{5}$, the resonance mode clearly disperses upward with increasing energy.

{\bf Data availability} The data that support the findings of this study are available from the corresponding author upon request.


\begin{thebibliography}{}
\bibitem{monthoux} Monthoux, P., Pines, D., and Lonzarich, G. G., Superconductivity without phonons. Nature {\bf 450}, 1177-1183 (2007).
\bibitem{scalapino} Scalapino, D. J., A common thread: The pairing interaction for unconventional
superconductors, Rev. Mod. Phys. {\bf 84}, 1383 (2012).
\bibitem{Keimer} Keimer, B., Kivelson, S. A., Norman, M. R., Uchida, S., and Zaanen, J.,
From quantum matter to high-temperature superconductivity in copper oxides, Nature {\bf 518}, 179-186 (2015).
\bibitem{Thompson} Thompson, J. D. and Fisk, Z., (2012), Progress in heavy-fermion superconductivity: Ce115 and related materials, J. Phys. Soc. Jpn. {\bf 81}, 011002 (2012).
\bibitem{maple} White, B. D., Thompson, J. D., Maple, M. B., Unconventional superconductivity in heavy-fermion compounds,
Physica C {\bf 514}, 246-278 (2015).
\bibitem{Mignod} Rossat-Mignod, J. {\it et al.}, Neutron scattering study of the YBa$_2$Cu$_3$O$_{6+x}$ system.
Physica C {\bf 185}, 86-92 (1991).
\bibitem{eschrig} Eschrig, M., The effect of collective spin-1 excitations on electronic spectra in
high-$T_c$   superconductors, Adv. Phys. {\bf 55}, 47-183 (2006).
\bibitem{Tranquada14} Tranquada, J. M., G. Xu, and Zaliznyak, I. A.,
Superconductivity, Antiferromagnetism, and Neutron Scattering, J. Mag.
Mag. Mater. {\bf 350}, 148-160 (2014).
\bibitem{Christianson} Christianson, A. D. {\it et al.}, Resonant Spin excitation in the high temperature
superconductor Ba$_{0.6}$K$_{0.4}$Fe$_2$As$_2$. Nature {\bf 456}, 930-932 (2008).
\bibitem{pcdai} Dai, P. C., Antiferromagnetic order and spin dynamics in iron-based superconductors,
Rev. Mod. Phys. {\bf 87}, 855 (2015).
\bibitem{Sato2001} Sato, N. K. {\it et al.}, Strong coupling between local moments and superconducting 'heavy' electrons in UPd$_2$Al$_3$, Nature {\bf 410}, 340-343 (2001).
\bibitem{CStock08} Stock, C., Broholm, C., Hudis, J., Kang, H. J., and Petrovic, C., Spin resonance in the $d$-Wave superconductor CeCoIn$_5$, Phys. Rev. Lett. {\bf 100}, 087001 (2008).
\bibitem{inosov11} Inosov, D. S. {\it et al.},
Crossover from weak to strong pairing in unconventional superconductors, Phys. Rev. B {\bf 83}, 214520 (2011).
\bibitem{GYu09} Yu, G., Li, Y., Motoyama, E. M., and Greven, M.,
A universal relationship between magnetic resonance and superconducting gap in unconventional superconductors,
Nat. Phys. {\bf 5}, 873-875 (2009).
\bibitem{Hirschfeld} Hirschfeld, P. J., Korshunov, M. M., and Mazin, I. I.,
Gap symmetry and structure of Fe-based superconductors, Rep. Prog. Phys. {\bf 74}, 124508 (2011).
\bibitem{dkmorr} Morr, D. K. and Pines, D., The resonance peak in cuprate superconductors, Phys. Rev. Lett. {\bf 81}, 1086 (1998).
\bibitem{chubukov08} Chubukov, A. V. and Gor'kov, L. P.,
Spin resonance in three-dimensional superconductors: the case of CeCoIn$_5$,
Phys. Rev. Lett. {\bf 101}, 147004 (2008).
\bibitem{Bourges} Bourges, P. {\it et al.}, The Spin Excitation Spectrum in Superconducting YBa$_2$Cu$_3$O$_6.85$, Science {\bf 288}, 1234-1237 (2000).
\bibitem{dai01} Dai, P. C., Mook, H. A., Hunt, R. D., and Do$\rm \breve{g}$an,
Evolution of the resonance and incommensurate spin fluctuations in superconducting YBa$_2$Cu$_3$O$_{6+x}$,
Phys. Rev. B {\bf 63}, 054525 (2001).
\bibitem{DReznik} Reznik, D. {\it et al.}, Dispersion of Magnetic Excitations in Optimally Doped Superconductor YBa$_2$Cu$_3$O$_{6.95}$, Phys. Rev. Lett. {\bf 93}, 207003 (2004).
\bibitem{Stock05} Stock, C. {\it et al.}, From incommensurate to dispersive spin-fluctuations: The high-energy inelastic spectrum in superconducting YBa$_2$Cu$_3$O$_{6.5}$, Phys. Rev. B {\bf 71}, 024522 (2005).
\bibitem{xylu13} Lu, X. Y. {\it et al.},
Avoided quantum criticality and magnetoelastic coupling in BaFe$_{2-x}$Ni$_x$As$_2$,
Phys. Rev. Lett. {\bf 110}, 257001 (2013).
\bibitem{MGKim13} Kim, M. G. {\it et al.}, Magnonlike dispersion of spin resonance in Ni-doped BaFe$_2$As$_2$, Phys. Rev. Lett. {\bf 110}, 177002 (2013).
\bibitem{Eremin05} Eremin, I., Morr, D. K., Chubukov, A.V., Bennemann, K. H., and Norman, M. R., Novel neutron resonance mode in $d_{x^2-y^2}$-wave superconductors, Phys. Rev. Lett. {\bf 94}, 147001 (2005).
\bibitem{MPAllan13} Allan, M. P. {\it et al.}, Imaging Cooper pairing of heavy fermions
in CeCoIn$_5$, Nat. Phys. {\bf 9}, 468-473 (2013).
\bibitem{BBZhou} Zhou, B. B. {\it et al.}, Visualizing nodal heavy fermion
superconductivity in CeCoIn$_5$, Nat. Phys. {\bf 9}, 474-479 (2013).
\bibitem{Eremin08} Eremin, I., Zwicknagl, G., Thalmeier, P., and Fulde, P., Feedback spin resonance in superconducting
CeCu$_2$Si$_2$ and CeCoIn$_5$, Phys. Rev. Lett. {\bf 101}, 187001 (2008).
\bibitem{Dyke14} Van Dyke, J., Massee, F., Allan, M. P., Davis, J. C., Petrovic, C., and Morr, D. K., Direct evidence for a magnetic f-electron–mediated pairing mechanism of heavy-fermion superconductivity in CeCoIn$_5$, PNAS {\bf 111}, 11663-11667 (2014).

\bibitem{Petrovic02} Petrovic, C., Bud'ko, S. L., Kogan, V. G., and Canfield, P. C.,
Effects of La substitution on the superconducting state of CeCoIn$_5$, Phys. Rev. B {\bf 66}, 054534 (2002).
\bibitem{Tanatar05} Tanatar, M. A. {\it et al.}, Unpaired electrons in the heavy-fermion superconductor CeCoIn$_5$,
Phys. Rev. Lett. {\bf 95}, 067002 (2005).
\bibitem{Panarin} Panarin, J., Raymond, S., Lapertot, G., Flouquet, J., and Mignot, J.-M., Effects of nonmagnetic La impurities on the spin resonance of Ce$_{1−x}$La$_x$CoIn$_5$ single crystals as seen via inelastic neutron scattering, Phys. Rev. B {\bf 84}, 052505 (2011).
\bibitem{RaymondS11} Raymond, S., Panarin, S., Lapertot, G., and Flouquet, J., Evolution of the spin resonance of CeCoIn$_5$ as a function of magnetic field and La substitution, J. Phys. Soc. Jpn. {\bf 80}, SB023 (2011).
\bibitem{Lshu} Shu, L. {\it et al.}, Correlated electron state in Ce$_{1-x}$Yb$_x$CoIn$_5$ stabilized by cooperative valence fluctuations, Phys. Rev. Lett. {\bf 106}, 156403 (2011).
\bibitem{Polyakov} Polyakov, A. {\it et al.},
Fermi-surface evolution in Yb-substituted CeCoIn$_5$, Phys. Rev. B {\bf 85}, 245119 (2012).
\bibitem{Dudy} Dudy, L., Denlinger, J. D., Shu, L., Janoschek, Allen, J. W., and Maple, M. B., Yb valence change in Ce$_{1-x}$Yb$_{x}$CoIn$_5$ from spectroscopy and bulk properties. Phys. Rev. B {\bf 88}, 165118 (2013).
\bibitem{Kim15} Kim, H. {\it et al.}, Nodal to nodeless superconducting energy-gap structure change concomitant
with Fermi-surface reconstruction in the heavy-fermion compound CeCoIn$_5$, Phys. Rev. Lett. {\bf 114}, 027003 (2015).
\bibitem{Erten} Erten, O., Flint, R., and Coleman, P., Molecular pairing and fully gapped superconductivity in Yb-doped CeCoIn$_5$, Phys. Rev. Lett. {\bf 114}, 027002 (2015).

\bibitem{PDas} Das, P. {\it et al.}, Magnitude of the magnetic exchange interaction in the heavy-fermion
antiferromagnet CeRhIn$_5$, Phys. Rev. Lett. {\bf 113}, 246404 (2014).
\bibitem{stock15} Stock, C., Rodriguez-Rivera, J. A., Schmalzl, K., Rodriguez, E. E., Stunault, A., and Petrovic, C.,
Single to multiquasiparticle excitations in the itinerant helical magnet CeRhIn$_5$, Phys. Rev. Lett. {\bf 114}, 247005 (2015).

\bibitem{smhayden} Hayden, S. M., Mook, H. A., Dai, P. C., Perring, T. G., and Do$\rm \breve{g}$an, F.,
The structure of the high-energy spin
excitations in a high-transition temperature superconductor, Nature {\bf 429}, 531-534 (2004).
\bibitem{Tranquada04} Tranquada, J. M. {\it et al.},
Quantum magnetic excitations
from stripes in copper oxide superconductors, Nature {\bf 429}, 534-538 (2004).
\bibitem{raymond15} Raymond, S. and Lapertot, G., Ising incommensurate spin resonance of CeCoIn$_5$:
a dynamical precursor of the $Q$ phase. Phys. Rev. Lett. {\bf 115}, 037001 (2015).
\bibitem{Kenzelmann08} Kenzelmann, M. {\it et al.}, Coupled superconducting and magnetic Order in CeCoIn$_5$,
Science {\bf 321}, 1652-1654 (2008).
\bibitem{Kenzelmann10} Kenzelmann, M. {\it et al.}, Evidence for a magnetically driven superconducting $Q$ phase of CeCoIn$_5$, Phys. Rev. Lett. {\bf 104}, 127001 (2010).
\bibitem{Gerber14} Gerber, S. {\it et al.}, Switching of magnetic domains reveals spatially inhomogeneous superconductivity, Nat. Phys. {\bf 10},
126-129 (2014).
\bibitem{Michal} Michal, V. P. and Mineev, V. P., Field-induced spin-exciton condensation in the $d_{x^2-y^2}$-wave superconductor CeCoIn$_5$,
Phys. Rev. B {\bf 84}, 052508 (2011).
\bibitem{Stock2012} Stock, C. {\it et al.}, it Magnetic Field Splitting of the Spin Resonance in CeCoIn$_5$, Pys. Rev. Lett. {\bf 109}, 167207 (2012).
\bibitem{Raymond2012} Raymond, S., Kaneko, K., Hiess, A., Steffens, P., and Lapertot, G., Evidence for three fluctuation channels in the spin resonance
of the unconventional superconductor CeCoIn$_5$, Phys. Rev. Lett. {\bf 109}, 237210 (2012).
\bibitem{Steffens} Steffens, P. {\it et al.},
Splitting of resonance excitations in nearly optimally doped Ba(Fe$_{0.94}$Co$_{0.06}$)$_2$As$_2$: an inelastic neutron scattering study with polarization analysis, Phys. Rev. Lett. {\bf 110}, 137001 (2013).
\bibitem{HQLuo13} Luo, H. Q. {\it et al.},
Spin excitation anisotropy as a probe of orbital ordering in the paramagnetic tetragonal phase of
superconducting BaFe$_{1.904}$Ni$_{0.096}$As$_2$,
Phys. Rev. Lett. {\bf 111}, 107006 (2013).
\bibitem{Stockert} Stockert, O. {\it et al.}, Magnetically driven superconductivity in CeCu$_2$Si$_2$, Nat. Phys., {\bf 7}, 119-124 (2011).
\bibitem{SJang} Jang, S. {\it et al.}, Resolution of the discrepancy between the variation of the physical properties of Ce$_{1-x}$Yb$_x$CoIn$_5$ single crystals and thin films with Yb composition, Philosophical Magazine {\bf 94}, 4219-4231 (2014).

\end{thebibliography}

\end{document}